\documentclass[prl,aps,twocolumn,floats,showpacs,superscriptaddress]{revtex4}

\usepackage{amsmath}
\usepackage{amssymb}
\usepackage{latexsym}

\usepackage{graphicx}
\usepackage{epsfig}

\usepackage{epic}
\usepackage{eepic}

\newcommand{\be}{\begin{equation}}
\newcommand{\ee}{\end{equation}}
\newcommand{\bea}{\begin{eqnarray}}
\newcommand{\eea}{\end{eqnarray}}

\newcommand{\rr}{\mathbf{r}}
\newcommand{\kk}{\mathbf{k}}
\newcommand{\ii}{\mathbf{i}}
\newcommand{\jj}{\mathbf{j}}
\newcommand{\llll}{\mathbf{l}}

\newcommand{\NN}{\mathcal{N}}
\newcommand{\DDr}{\Delta^{rnd}}
\newcommand{\al}{\alpha}
\newcommand{\oo}{\mathbf{0}}

\begin{document}
\title{Diffusion Processes on Power-Law Small-World Networks}

\author{Bal\'azs Kozma}
%\email{kozmab@rpi.edu}
\affiliation{Department of Physics, Applied
Physics, and Astronomy, Rensselaer Polytechnic Institute, 110
8$^{th}$ Street, Troy, NY 12180--3590, USA}

\author{Matthew B. Hastings}
%\email{hastings@lanl.gov}
\affiliation{Center for Non-linear Studies and Theoretical Division,
Los Alamos National Laboratory, Los Alamos, New Mexico 87545, USA}

\author{G. Korniss}
%\email{korniss@rpi.edu}
\affiliation{Department of Physics, Applied
Physics, and Astronomy, Rensselaer Polytechnic Institute, 110
8$^{th}$ Street, Troy, NY 12180--3590, USA}

\begin{abstract}
  We consider diffusion processes on power-law small-world networks in
  different dimensions.  In one dimension, we find a rich phase diagram,
  with different transient and recurrent phases, including a critical line
  with continuously varying exponents. The results were obtained using
  self-consistent perturbation theory and can also be understood in terms
  of a scaling theory, which provides a general framework for understanding
  processes on small-world networks with different distributions of
  long-range links.
\end{abstract}

\date{\today}
\maketitle

Studying diffusion-related processes on networks, one can gain insight
into synchronization \cite{MZK05} and spreading phenomena in natural, artificial, and
social systems \cite{NEWMAN05}. 
In this Letter, we consider a class of models on small-world (SW) networks \cite{Watts} with a
power-law probability distribution of long-range links, where the
probability of two nodes being connected goes as $r^{-\alpha}$, with
some exponent $\alpha$ \cite{kleinberg,JB00,sh_path,PR05}.
Such a structure can emerge in
synchronization problems of distributed computing \cite{Korniss}
where synchronization is achieved by introducing random communications
between distant processors.  Choosing a power-law SW network may be
preferable, as it lowers the cost associated with communications. 
It was also recently argued \cite{PR05} that ``wiring-cost'' considerations for
spatially embedded networks, such as cortical networks \cite{LS03} or
on-chip logic networks \cite{DDM98}, can generate such
power-law-suppressed link-length distribution.  
A
physical example of such a power-law SW network arises in diffusion on 
a randomly folded polymer 
discussed below \cite{SMB97,BG2003}.

The addition of random long-range links to a regular $d$-dimensional
network, producing a SW network, leads in many
cases to a crossover to mean-field-like behavior \cite{mft},
effectively becoming equivalent to averaging over the long-range links
in an annealed fashion. Here we focus on systems where it is not the
case, and the contrast between the quenched and annealed systems is
strong \cite{JB00,khk}. We develop a perturbative method for the quenched
network, which is asymptotically exact, as confirmed by numerics.

Varying the exponent $\alpha$, controlling the distribution of length of the links, the diffusive dynamics
gives rise to a rich phase diagram, as the connection topology interpolates
between the original ``plain'' SW ($\alpha$$=$$0$) and the purely short-range connected
($\alpha$$=$$\infty$) network.
The phase diagram can also be understood in terms of simple scaling
ideas, showing that the breakdown of mean-field theory is associated with
the relevance of the operator for scattering off a single link.

{\it Diffusion-Related Processes---} We start with a
$d$-dimensional lattice of linear size $L$ and add long-range
links, connecting two sites with a probability proportional to $p
r^{-\alpha}$, where $r$ is the distance between the two sites, $p$
is the probability of a site to have a random link, and $\alpha$
is the exponent of the decay, ranging from $0$ to $\infty$.
Although the original problem can have only one random link
between two distant sites, the analytics are slightly simpler if
pairs of sites may be connected with multiple links with a Poisson
distribution of links; both cases have the same universal behavior
in the small-$p$ limit.  Thus, in this work, we set the
probability, that sites $i$ and $j$ are connected by $n$ random
links, equal to $\frac{\left(p/(\NN
    |i-j|^{\alpha})\right)^n}{n!} e^{-p/(\NN |i-j|^{\alpha})}$, where
$\NN=\sum_{j \ne 0} |j|^{-\alpha}$.

One case in which the diffusion equation arises is a macromolecule randomly
moving along a polymer chain jumping over adjacent segments with nonzero
probability \cite{SMB97}; if the macromolecule motion is fast compared to rearrangements
of the links, then the network may be considered as quenched \cite{SMB97,MONA99}, 
while if the macromolecule motion is slow compared to the link rearrangements 
then the network is annealed \cite{SMB97,BG2003}. 
The equation describing random walk processes is
\begin{equation}
  \partial_t  P_{\rr'}(t)=-\sum_{\rr''}   \left(  \Delta_{\rr'  \rr''}+q
    \Delta^{rnd}_{\rr' \rr''} \right) P_{\rr''}(t),
\end{equation}
where $P_{\rr'}(t)$ is the probability of finding the walker at site $\rr'$
at time $t$, $-\Delta_{\rr' \rr''}$ is the discretized version of the
$d$-dimensional diffusion (or Laplace) operator, and $-\Delta^{rnd}_{\rr'
  \rr''}$ is the diffusion operator on the random links with
diffusion coefficient $q$ i.e.  in one
dimension $\Delta_{ij}= 2\delta_{i,j} -\delta_{i,j-1} -  \delta_{i,j+1}$
and $\Delta^{rnd}_{ij}=\delta_{ij}\sum_{l \ne i} J_{il} -J_{ij}$, where
$J_{ij}=n$ if there are $n$ links connecting sites $i$ and $j$.
The Green's function,
$G_{\rr',\rr''}(t)$, of this model is the solution of the equation with
initial condition $P_{\rr'}(t=0)=\delta_{\rr',\rr''}$. Because the net
probability has to be conserved during the process, the transition matrix
has a zero eigenmode so the matrix is invertible only in the subspace
orthogonal to this mode. Consequently, the limiting case of the temporally
Fourier-transformed Green's function,
$G_{\rr',\rr''} = \lim_{\omega\to 0}\left(i\omega + \Delta + q \Delta^{rnd}\right)_{\rr',\rr''}^{-1}$,
is only invertible in the same subspace. $G_{\rr',\rr'}$, in the
context of the random walk, is related to the return probability and
its scaling behavior with $L$ will determine whether the walk is
transient or recurrent. Earlier numerical work \cite{JB00} 
for $d$$=$$1$ suggested $\alpha \approx 2$
as a crossover point from the transient to the recurrent phase.
Here we compute the scaling properties of the propagator analytically
(and confirm it numerically), 
and precisely construct the full phase diagram for all $d$.

The second case, relevant to synchronization problems on SW networks in
a noisy environment, is the Edwards-Wilkinson process \cite{khk}
\begin{equation}
\partial_t h_{\rr'}(t)=-\sum_{\rr''} \left( \Delta_{\rr' \rr''}+q
  \Delta^{rnd}_{\rr' \rr''} \right) h_{\rr''}(t) + \eta_{\rr'}(t),
\label{EW_eq}
\end{equation}
where $h$ is the surface height and $\eta$ is a delta-correlated Gaussian
white noise with variance 2. In this case, the Green's function is the
steady-state two-point correlation function between $h_{\rr'}$ and $h_{\rr''}$,
so $G_{\rr',\rr''} = \langle h_{\rr'}
h_{\rr''} \rangle = \left( \Delta + q \Delta^{rnd}
\right)_{\rr',\rr''}^{-1}$,
and again, the function is defined in the space
orthogonal to the uniform zero mode. In the context of the
Edwards-Wilkinson equation, $G_{\rr',\rr'}$ is the measure of the
surface roughness or width \cite{khk}.

Ultimately, we are interested in the average of these quantities over all
realizations of the networks. This averaging (denoted by $[\ ]$)
restores translation invariance for $[G_{\rr',\rr''}]$ and therefore
its spatial behavior will depend only on the difference
$\rr''-\rr'$. Hence, we define
\begin{eqnarray}
G(\rr)\equiv [(\Delta+q\DDr)^{-1}]_{\rr',\rr'+\rr},
\label{eq:G(r)}
\end{eqnarray}
where $\rr$ is the spatial separation of the two sites. We expect
that when the probability of the random links decays rapidly
enough ($\alpha$$\to$$\infty$), the large scale behavior of the
system is not affected by these links.  When the probability of
the random links decays slowly ($\alpha$$\to$$0$), the results of the
original SW network should be recovered \cite{khk}.  In between,
there is a transition which is the focus of this Letter.  Our
starting point is an annealed calculation that turns out to be the
first order result of a naive perturbation expansion for the
quenched system.  For small $p$ this expansion breaks down and we
apply a self-consistent calculation.

%%%%%%%%%%%%%%%%%%%%%%%%%%%%%%%%%%%%%%%%%%%%%%%%%%%%%%%%%%%%%%%%%%%%%%
\begin{figure}[t]
\begin{center}
\epsfxsize8cm \epsfbox{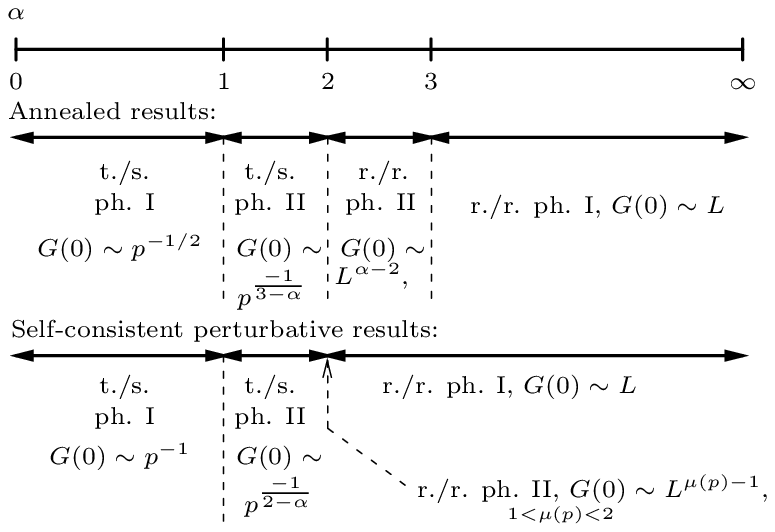}
\end{center}
\vspace*{-0.75cm}
\caption{Phase diagram for one dimension. Results from the
    annealed calculation (top) and from the self-consistent perturbative
    calculation for the quenched case (bottom)
    showing different behavior in the transient/smooth phases
    (t./s. ph.) and in the recurrent/rough (r./r.) ones.}
\label{fig:1d-phases}
\end{figure}
%%%%%%%%%%%%%%%%%%%%%%%%%%%%%%%%%%%%%%%%%%%%%%%%%%%%%%%%%%%%%%%%%%%%%%

{\it Annealed Calculation---} First, we consider the annealed predictions.
In this calculation, to approximate the behavior of $G(\rr)$, the random
links are replaced with their expected number, i.e. the random interactions
are annealed \cite{JB00,SMB97,BG2003,racz}. That is, in (\ref{eq:G(r)}),
$\Delta^{rnd}$ is substituted with % $[\Delta^{rnd}]$.
$[\Delta^{rnd}]$. $[\Delta^{rnd}]_{\rr,\rr}=p$, and $
[\Delta^{rnd}]_{\rr,\rr'} = \frac{-p}{\NN} \frac{1} {\
  \left|\rr-\rr'\right|^{\alpha}}$ for $\rr \ne \rr'$.  Because of its
translational invariance, this operator is diagonal in Fourier-space,
likewise $\Delta$.  In the large system-size limit, its eigenvalues in
Fourier basis are:
  $[\DDr](\kk)\propto p$ for $0<\alpha<d$;
  $[\DDr](\kk)\propto p k^{\alpha-d}$ for $d<\alpha<d+2$; and
  $[\DDr](\kk)\propto p k^2+...$ for $d+2<\alpha$.
Hence, $G(\rr)$ can be
obtained by inverse Fourier-transformation and
\begin{equation}
  G(\oo) = \frac{1}{L^d}\sum_{\kk \ne 0}
    \frac{1}{k^2+q\ [\DDr](k)}
  \label{eq:MF}
\end{equation}
To calculate the scaling properties of this quantity, with respect to $p$,
$q$, and $L$, the sum can be approximated with its integral-form limit.
The results for one dimension are sketched in Fig.~\ref{fig:1d-phases} (top).
For $\al<2$, the integral (\ref{eq:MF}) converges at both ends, so is
system size independent for large $L$ and diverges as $p \to 0$. There are
two regimes, distinguished by their scaling behavior (reflecting
the properties of the underlying random walk/surface) :
(1) {\it transient/smooth phase I}, where the system behaves as if the
long range links were uniformly distributed, and $G(0)$ is $\al$-independent;
(2) {\it transient/smooth phase II}, where $G(0)$ is still finite in the
thermodynamic limit, but has an $\alpha$-dependent divergence as $p \to 0$.
For $\al>2$, $G(0)$ has an infrared divergence resulting in two system-size
dependent regimes:
(1) {\it recurrent/rough phase I}, where the system behaves as if there were
no long range links, and $G(0)$ diverges linearly with the
system-size, $L$;
(2) {\it recurrent/rough phase II}, with sublinear scaling with respect to
the system-size. At the boundaries of these phases, logarithmic
corrections are present.

{\it Self-consistent Perturbative Calculation for the Quenched Network---}
Following \cite{khk}, we calculate $G(\mathbf{0})$ perturbatively, treating
$\DDr$ as a weak scattering potential.
Expanding Eq.(\ref{eq:G(r)}), $G=G^0-G^0[q \DDr] G^0+G^0[ q
\DDr G^0 q \DDr] G^0 - \dots$, where $G^0=\Delta^{-1}$, the Green's
function with no random interactions introduced. Rearranging the formula,
one obtains the Dyson equation,
$G=G^0-G^0 \Sigma G = (\Delta+\Sigma)^{-1}$,
where $\Sigma=q[\DDr]-q^2[\DDr G \DDr]_c + \dots$ is the self-energy and
$[\ ]_c$ represents the cumulants of the random variables
$\DDr_{\rr,\rr'}$, for example $[\DDr_{\ii \jj} \DDr_{\kk \llll}]_{c} =
[\DDr_{\ii \jj} \DDr_{\kk \llll}]-[\DDr_{\ii \jj}] [\DDr_{\kk \llll}]$ and
so on.  Solving the equation to first order, $\Sigma=q[\DDr]$, the annealed
result is recovered. Substituting this first-order result into the next
order term of $\Sigma$, $-q^2[\DDr G \DDr]_c$, it can be shown that the
correction becomes dominant over the first-order term as $p \to 0$, and so
the perturbation expansion breaks down.

To have an operative perturbation expansion, as a first step $\Sigma$ has
to be calculated involving all the scatterings from a single link, then
corrections representing scatterings from multiple links may be
successively obtained.  The closed form of $\Sigma$ for the single-link
problem is
\begin{equation}
\Sigma^{sl}({\rr \ne \mathbf{0}})=-q\frac{p}
{\NN r^{\alpha}} \frac{1} {1+2q( G(\mathbf{0})-G(\rr))}
\end{equation}
and $\Sigma^{sl}(\mathbf{0})= - \sum_{\rr \ne \mathbf{0}}
\Sigma^{sl}(\rr)$.  Since $G$ and $\Sigma^{sl}$ are coupled, they
have to be solved for {\em self-consistently}.  To do this, we make an
{\em ansatz} for $\Sigma^{sl}$, namely, in Fourier-space, $\Sigma^{sl}(k)=s
k^{\mu}$ for small $k$-s, $s$ and $\mu$ to be determined.  Using this form
of $\Sigma^{sl}$, the Green's function, in Fourier space,
becomes $G(k)= (k^2+sk^{\mu})^{-1}$. Its long-distance real-space behavior,
as in the annealed case, can be calculated with the integral-form limit of
the expressions and $\Sigma^{sl}$ should be treated similarly.

For one dimension, as $p \to 0$ at non-vanishing $q$, one finds several
cases.  For
$\alpha<1$, $\mu=0, s=p^2, G(0)\propto p^{-1}$.  For
$1<\alpha<2$, $\mu=\alpha-1, s= p^{\frac{3-\alpha}{2-\alpha}}, 
G(0) \propto p^{\frac{-1}{2-\alpha} }$.  For
$\alpha>2$,
$\mu=\alpha,s=p, G(0) \propto L$,
while for $\alpha$$=$$2$, $\mu$ is given by the solution of the self-consistent
formula: $p \frac{3}{2\pi^2} \frac{ \cos(\frac{\mu\pi}{2})\Gamma(-\mu)}{
  \sin(\frac{\mu\pi}{2})\Gamma(1-\mu) }{=1}$, $1$$<$$\mu$$<$$2$.
%%%%%%%%%%%%%%%%%%%%%%%%%%%%%%%%%%%%%%%%%%%%%%%%%%%%%%%%%%%%%%%%%%%%%%%%%%%%%%%%
The leading order results are sketched in 
Fig.~\ref{fig:1d-phases} (bottom). Similar phases appear as were predicted by the
annealed argument, but in {transient/smooth phase II}, we have
a different scaling property, while
{recurrent/rough phase I}
spans a wider interval of the $\alpha$-axis. Finally, {
  recurrent/rough phase II} is {\em collapsed} to one point on this
axis, and in this phase, the sublinear behavior of $G(0)$ is no longer
determined by the distance-distribution of the random links but rather
by their density, $p$, through the self-consistent formula. The form
of this equation is approximate, but the fact that the exponent $\mu$
depends continuously on $p$ is likely to be correct, in light of the
scaling arguments below.
%%%%%%%%%%%%%%%%%%%%%%%%%%%%%%%%%%%%%%%%%%%%%%%%%%%%%%%%%%%%%%%%%%%%%%
\begin{figure}[t]
\begin{center}
\epsfxsize8cm \epsfbox{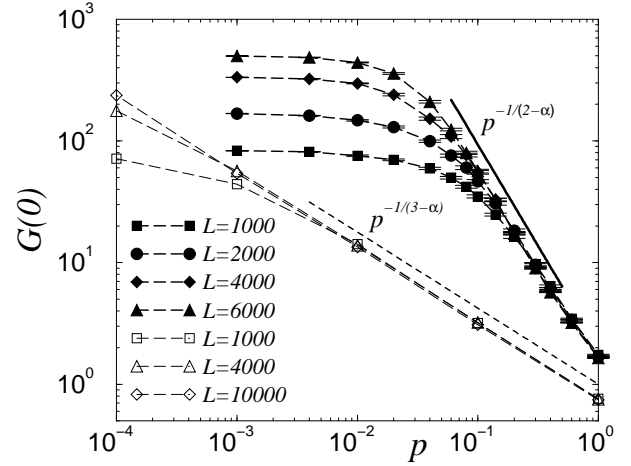}
\end{center}
\vspace*{-0.75cm}
\caption{
\label{fig:a_1.4}
Finite-size propagators $G(0)$ vs.\ $p$ for $\alpha$$=$$1.4$, $q$$=$$1$
(transient/smooth phase II) for various system sizes.
Solid symbols represent data obtained by exact numerical
diagonalization averaged over 100 realizations for the quenched network. The
solid straight line represents the slope obtained from the perturbative analytic
calculations in the $L$$\to$$\infty$ limit.
Open symbols correspond to the steady-state averaged propagator on the
annealed network with the straight dashed line being the mean-field
analytic result in the $L$$\to$$\infty$ limit.}
\end{figure}
%%%%%%%%%%%%%%%%%%%%%%%%%%%%%%%%%%%%%%%%%%%%%%%%%%%%%%%%%%%%%%%%%%%%%%%%

{\it Scaling---} These results for the quenched network can be understood
via a simple scaling argument.  The Green's function has dimension
$l^{2-d}$, where $l$ is some length scale, as can be
seen by considering the Green's function
$G^0(r)=\int {\rm d}^d k \exp(i k r)/k^2$.
The dimension of $q$ is
$l^{d-2}$, as $qG$ is dimensionless.

The dimension of $p$ is $l^{-d}$ for $\alpha\leq d$ and $l^{\alpha-2d}$ for
$\alpha\geq d$.  This arises from perturbation theory, but can also be
understood geometrically: the probability that
a block of linear size $l$ has a long-range link with a length
at least $l$ is $pl^d$ for $\alpha\leq d$
and $pl^{2d-\alpha}$ for $\alpha\geq d$.
Thus, at a length scale $l$, for
$\alpha\geq d$, the important dimensionless quantities are:
$\tilde q=q l^{2-d}$,
$\tilde p=p l^{2d-\alpha}$, and
$\tilde \omega=\omega l^2$.
There are several cases in the limit of small $p$ and non-vanishing $q$.
If $2d<\alpha$, then $\tilde p$ scales to zero, and there are no links
at large length scales; the system looks like a network without long-range
links, giving $G(0)\propto L$ as in recurrent/rough phase I.

If instead $2d>\alpha>d$, then at some scale $l\propto p^{-1/(2d-\alpha)}$,
the density of long-range links becomes of order unity.  This case breaks
into two further subcases.  If $d<2$, then $q$ is relevant; if $p$ is small
and $q$ is non-vanishing, then higher order corrections will cut off the
growth in $\tilde q$, so $\tilde q$ at scale $l$ will be of order unity, and
will
be independent of $q$ at lattice scale.  This tells us that the only
meaningful dimensional number is $p$ (since $q$ is not important at this
scale), and dimensional analysis gives $G(0)\propto p^{-(2-d)/(2d-\alpha)}$ as
in transient/smooth phase II.
This is a case in which mean-field theory breaks
down because there is typically only one link at the scale $l\propto
p^{-1/(2d-\alpha)}$ and the
scattering off links is strong.  If instead $d>2$, then $\tilde q$
decreases with increasing $l$,
and at $l\propto p^{-1/(2d-\alpha)}$ the system has a high
density of weak links where mean-field theory is valid.  Near $d=2$, an
$\epsilon$-expansion is available \cite{ee} to compute higher corrections
and determine prefactors in $G(0)$.  
Finally, for $\alpha=2d$, $\tilde p$
is marginal, giving the continuously varying exponent of
recurrent/rough phase II.

In the annealed case, one has a single number $pq$, and the
corresponding dimensionless coupling constant is 
$\tilde{pq} =pq l^{d+2-\alpha}$. This becomes relevant for 
$\alpha<d+2$, and, e.g., yields the scaling $G(0)\propto (pq)^{-(2-d)/(d+2-\alpha)}$ 
in the transient/smooth phase II.

%%%%%%%%%%%%%%%%%%%%%%%%%%%%%%%%%%%%%%%%%%%%%%%%%%%%%%%%%%%%%%%%%%%%%%%
\begin{figure}[t]
\begin{center}
\epsfxsize8cm
\epsfbox{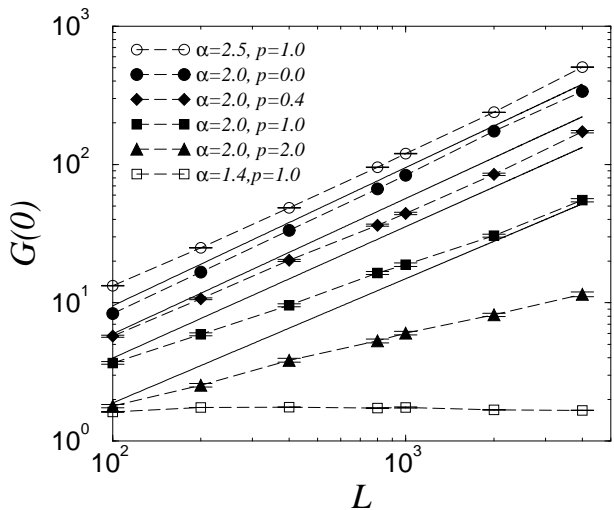}
\end{center}
\vspace*{-0.75cm}
\caption{
\label{fig:a_2.0}
$G(0)$ vs the system size $L$ (all data for $q$$=$$1$) from
exact numerical diagonalization.
Data in the recurrent/rough phase I ({\large$\circ$} symbols)
indicate the linear scaling with $L$ (data is shifted up for clarity).
Filled symbols represent the scaling behavior in the recurrent/rough
phase II ($\alpha$$=$$2$) for various values of $p$, indicating the
$p$-dependence of the exponent. The solid lines, representing the
slopes obtained from the self-consistent formula for the same set of
values of $p$,
show significant deviations from the corresponding data for larger
values of $p$.
Data in the transient/smooth phase II ($\Box$ symbols),
indicating a finite $G(0)$ in the $L$$\to$$\infty$ limit, are
also shown for comparison.}
\end{figure}
%%%%%%%%%%%%%%%%%%%%%%%%%%%%%%%%%%%%%%%%%%%%%%%%%%%%%%%%%%%%%%%%%%%%%%%%

{\it Numerical Results---} For the annealed case, we numerically
integrated Eq.~(\ref{EW_eq}) where the underlying power-law SW network
was also reconfigured at every timestep, and measured the
width $G(0)$ in the steady state.
The results in Fig.~\ref{fig:a_1.4}
indicate that the mean-field approximation
is exact for the annealed system. For the quenched case, we compared our
analytic results to those obtained from exact numerical
diagonalization of the discretized Laplacian for
representative points in the phase diagram. In
Fig.~\ref{fig:a_1.4} we show the main features of transient/smooth
phase II ($\alpha$$=$$1.4$). For a region of $p$ values, good
agreement between the two results can be observed. For finite
systems of size $L$, the crossover in $G(0)$ (see below) occurs
when $p<p_\times \propto L^{\alpha-2}$. For large values of $p$,
deviation can be observed due to possible higher order corrections
in $p$. In Fig.~\ref{fig:a_2.0} we show the $L$-dependence of
$G(0)$ in the recurrent/rough phases. While in the recurrent/rough
phase I it diverges as $G(0)\propto L$, in the somewhat
``degenerate'' recurrent/rough phase II (collapsed onto a single
point, $\alpha$$=$$2$) the exponent of the divergence depends
continuously on $p$, $G(0)\propto L^{\mu(p)-1}$. Although the
self-consistent formula (solved numerically) predicts this latter
feature qualitatively, the actual exponent appears to be affected
by higher order corrections.

{\it Crossovers---} There are two competing terms determining the large
scale behavior of the Green's function, the diffusion through the regular
links, $\Delta$, and through the random ones, $\DDr$.  For small system sizes,
the number of the random links is small, their contribution to the
ensemble average is negligible, the behavior is close to that of a
$d$-dimensional unperturbed system. The effect of the random interactions
takes over when the self-energy generated by these random links start to
dominate the infrared behavior of $G(k)$, which is when
$k_{\times}^2 \approx \Sigma(k_{\times})$.
Thus, for the quenched case, in the {transient/smooth phase II} in
one dimension, for $L \ll L_{\times} = k_{\times}^{-1} \propto p^{\frac{-1}{2-\alpha}}$,
$G(0)\propto L$; for $L \gg L_{\times}$, $G(0)\propto
p^{\frac{-1}{2-\alpha}}$. For the annealed system, the corresponding
crossover occurs at $L_{\times} \propto
p^{\frac{-1}{3-\alpha}}$, leading to $G(0)\propto
p^{\frac{-1}{3-\alpha}}$ for $L \gg L_{\times}$.

{\it Discussion---} We investigated diffusion related processes on SW
networks with distance dependent distribution of the random links. The
results are summarized in the phase diagram of Fig.~1, with dramatically
different behavior between the quenched and annealed case.  The results for
the quenched network can be understood via simple scaling arguments, based
on the probability of finding links at a given length scale.
We expect that these scaling arguments can be generalized
to systems with an arbitrary probability $p(r)$ of having long-range links
and thus will be applicable to a wide-range of networks
in which mean-field theory breaks down.

{\it Acknowledgments---} This work was supported by NSF Grants
DMR-0113049, DMR-0426488, the Research Corporation, and by US DOE contract
W-7405-ENG-36. BK also acknowledges support from the LANL Summer Student
Program.

%%%%%%%%%%%%%%%%%%%%%%%%%%%%%%%%%%%%%%%%%%%%%%%%%%%%%%%%%%%%%%%%%%%%%%%%%

\end{document}